\documentclass[11pt,twoside]{article}
\usepackage{asp2014}

\aspSuppressVolSlug
\resetcounters

\begin{document}

\title{Non-parametric stellar LOSVD analysis}

\author{Damir~Gasymov$^{1,2}$ and Ivan~Katkov$^{1,3,4}$}
\affil{$^1$Sternberg Astronomical Institute, Moscow State University, Moscow, Russia;}
\affil{$^2$Department of Physics, Moscow State University, Moscow, Russia;}
\affil{$^3$New York University Abu Dhabi, Abu Dhabi, UAE;}
\affil{$^4$Center for Astro, Particle, and Planetary Physics, NYU Abu Dhabi, Abu Dhabi, UAE.}

\paperauthor{Ivan~Katkov}{ivan.katkov@nyu.edu}{0000-0002-6425-6879}{New York University Abu Dhabi}{Center for Astro, Particle, and Planetary Physics}{Abu Dhabi}{Abu Dhabi}{129188}{United Arab Emirates}
\paperauthor{Damir~Gasymov}{gasymov.df18@physics.msu.ru}{0000-0002-1750-2096}{M.V.~Lomonosov Moscow State University}{Sternberg Astronomical Institute}{Moscow}{}{119992}{Russia}




  
\begin{abstract}

Ill-posed inverse problems are common in astronomy, and their solutions are unstable with respect to noise in the data. Solutions of such problems are typically found using two classes of methods: parametrization and fitting the data against some predefined function or a solution with a non-parametrical function using regularization. Here we are focusing on the latter non-parametric approach applied for the recovery of complex stellar line-of-sight velocity distribution (LOSVD) from the observed galaxy spectra. Development of such an approach is crucial for galaxies hosting multiple kinematically misaligned stellar components, such as 2 stellar counter-rotating disks, thin and thick disks, kinematically decoupled cores, and others. 

Stellar LOSVD recovery from the observed galaxy spectra is equivalent to a deconvolution and can be solved as a linear inverse problem. To overcome its ill-posed nature we apply smoothing regularization. Searching for an optimal degree of smoothing regularization is a challenging part of this approach. Here we present a non-parametric fitting technique, discuss its potential caveats, perform numerous tests based on synthetic mock spectra, and show real-world application to MaNGA spectral data cubes and some long-slit spectra of stellar counter-rotating galaxies. 
  
\end{abstract}

\section{Introduction}

One of the common inverse problems in astronomy is the deconvolution problem, which arises in a large number of situations.
Of particular interest is in the field of galaxy physics, since the spectra of observed galaxies can be considered as a convolution of the spectrum of the stellar population with the line-of-sight velocity distribution (LOSVD).
Recovery of stellar LOSVDs allows us to study the kinematical and dynamical properties of galaxies.
The standard way to extract kinematic information nowadays is to use Gauss-Hermite parameterization \citep{vanderMarel1993ApJ...407..525V} for LOSVD during full spectral fitting with \textsc{ppxf} \citep{Cappellari2017MNRAS.466..798C}, \textsc{NBursts} \citep{Chilingarian2007IAUS..241..175C} or similar tools.
However, the Gauss-Hermite (GH) parameterization is not always applicable; for example, the two-peaked LOSVD structure of counter-rotating galaxies cannot be described by GH shape (see Fig.~3 in \citet{Katkov2013ApJ...769..105K}).
We began to develop and apply a nonparametric approach for LOSVD recovery as part of our previous studies of counter-rotating galaxies \citep{Katkov2013ApJ...769..105K, 2016MNRAS.461.2068K} and thick/thin disk decomposition of edge-on disk galaxies \citep{2020MNRAS.493.5464K}.
There are other studies that use non-parametric stellar LOSVD recovery \citep{2021A&A...646A..31F, 2006MNRAS.365...74O, 1992MNRAS.254..389R}.
Now we are further improving our approach, perform tests based on synthetic mock spectra, discuss possible shortcomings, and demonstrate application on MaNGA spectral data and long slit spectra of a counter-rotating galaxy.

\section{Method description}
\begin{figure}
    \centering 
    \includegraphics[width=0.65\columnwidth,trim={0cm 1.8cm 0cm 0.3cm},clip]{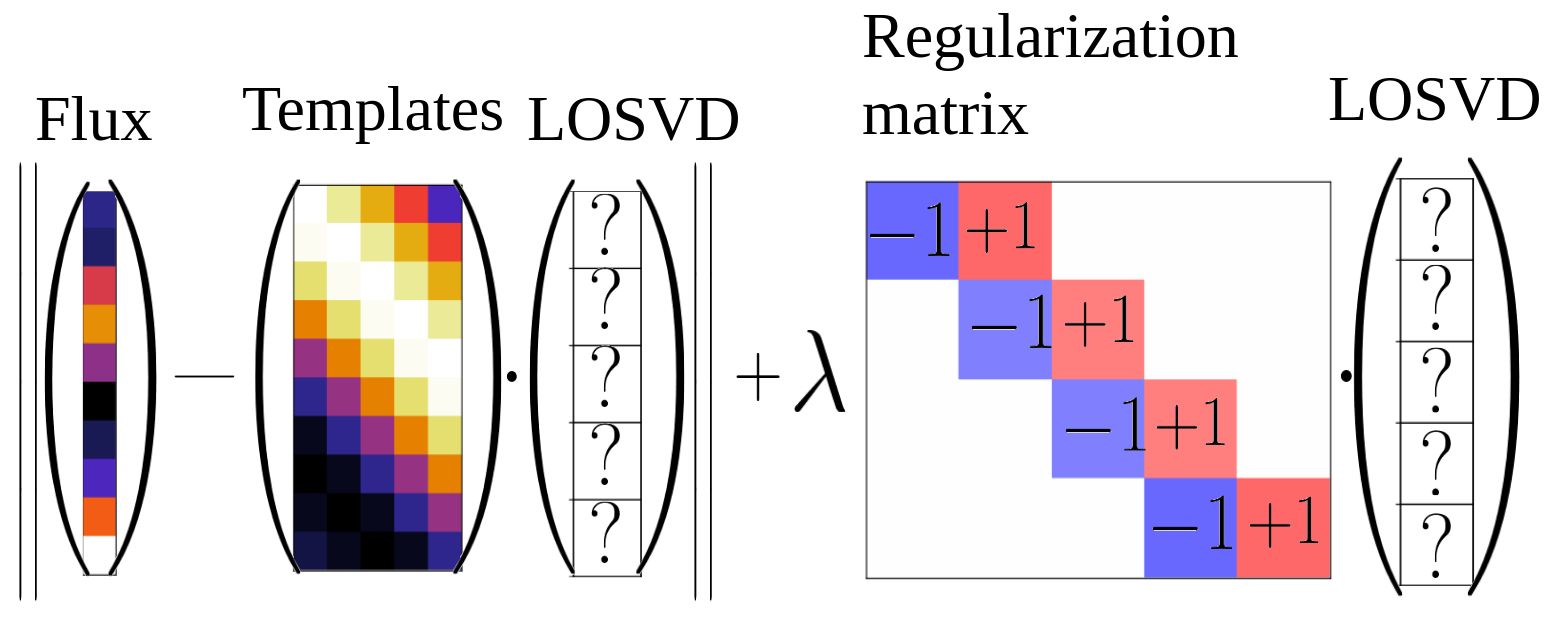}
    \caption{Graphical representation of the deconvolution problem using 1st order smoothing regularization. ``Flux'' is a vector of logrebinned observed galaxy spectrum, ``templates'' is a matrix of shifted stellar population template according to a given velocity offset. LOSVD - unknown solution of the problem.}
    \label{fig:image}
\end{figure}

The logarithmically rebinned spectrum of galaxy $Y$ can be considered as a convolution of the stellar population template with the stellar LOSVD $\mathcal{L}$ as the convolution kernel.
Convolution can be expressed as a linear problem $A \cdot \mathcal{L} = Y,$ where $A$ is a matrix, each $i$-th column of which contains a stellar population template shifted according to the velocity offset for a given $i$-th element of LOSVD.
The solution to the problem can be found using the linear least-squares method.
However, the solution is too sensitive to noise in the observed spectrum, so we obliged to use regularization in the following shape:
$$||Y - A\cdot \mathcal{L}|| + \lambda ||R \cdot \mathcal{L}|| \xrightarrow[\mathcal{L}]{} min,$$
where the regularization term imposes additional restrictions on the desired solution.
The regularization matrix $R$ can be of different types depending on our needs, spectrum quality and expected shape of LOSVD.
We have three types of regularization: \textsc{L2}, where $R$ is the identity matrix, which introduces a penalty for large values of LOSVD;
\textsc{smooth1} - first-order smoothing regularization, where $R$ represents a first-order difference matrix and contains [-1,1] values on the matrix diagonal (as shown schematically in Fig.~\ref{fig:image}), the regularization tries to minimize the first derivative of LOSVD producing smooth solution; \textsc{smooth2} - second-order smoothing regularization, matrix $R$ contains [-1,2,-1] to use difference approximation of the second derivative.
Optionally, we also have a regularization to minimize LOSVD values at high velocity amplitudes, the so-called \textit{wing} regularization.
In this case, $R$ is a diagonal matrix whose diagonal elements are $\sim V^2$ or a similar function, where the penalty coefficients must grow from the center to the edge of the LOSVD.

Recently, we added an iterative fitting, where in the first iteration the algorithm finds corrections to the initial Gauss-Hermite LOSVD guess, and in the subsequent iterations it looks for corrections to the previous full non-parametric solution.
Smoothing and other regularizations are imposed on the corrections, not on the full solution.
We use the \textsc{nbursts} fitting results as parameters of an initial GH approximation as well as stellar population template.

Spectra from nearby spatial bins should have similar LOSVDs, since we assumed gradual changing of kinematic properties and/or the effect of atmospheric seeing.
Therefore, we extended the smoothing regularization functionality to act along the spatial axis. To date, this feature works only for long-slit spectra.

We have made the Python implementation of our method publicly available on pypi \url{https://pypi.org/project/sla/} and Github repo \url{https://github.com/gasymovdf/sla}.

\section{Tests on synthetic spectra}

\begin{figure}
    \centering 
    \includegraphics[width=0.7\columnwidth,trim={0cm 3.5cm 0cm 1.7cm},clip]{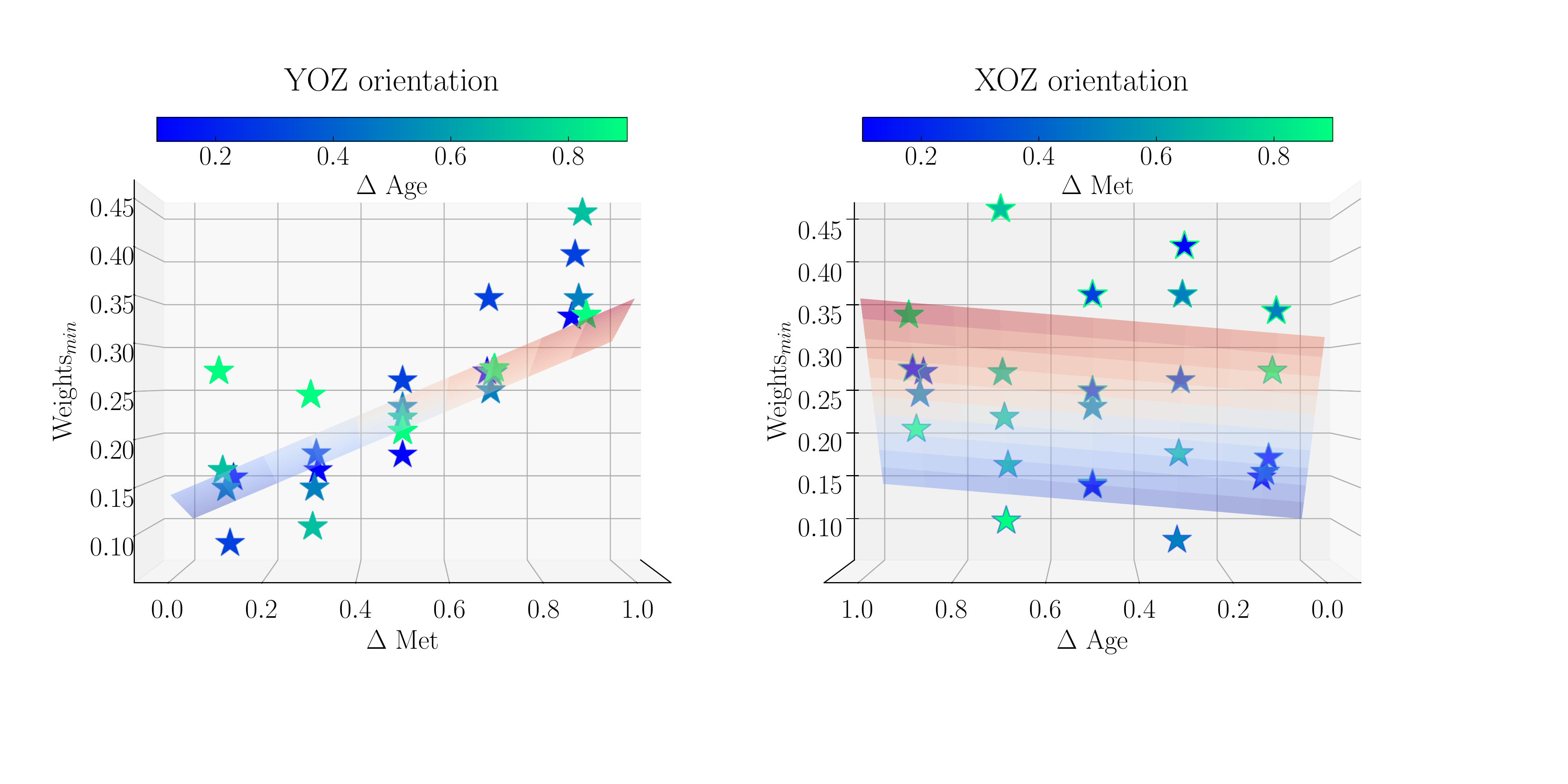} 
    \caption{Result of analysis of minimal detectable contribution of secondary component to the integral light (W$_{min}$) in dependence on difference in age and metallicity of the main and secondary components.}
    \label{fig:age_met_weights}
\end{figure}

To test our method, especially for the study of counter-rotating galaxies, we created 2,000,000 synthetic spectra with spectral resolution R=2000 in the spectral range 4200-5500~\AA\ with a signal-to-noise ratio of 30.
Each spectrum is the sum of two Simple Stellar Population (SSP) broadened with an individual Gaussian LOSVD and described by various parameters of stellar populations (Age$_1$, Age$_2$, Met$_1$, Met$_2$ -- the age and metallicity) and kinematic parameters (V$_1$ = 0, V$_2$ -- the velocity of the second component relative to the first, $\sigma_1 = \sigma_2$ is the approximation of equal velocity dispersions and W$_1$ = 1 - W$_2$ and W$_2$ are the weights of the corresponding components).

One of the goals of synthetic tests is to find the minimum detectable contribution of the secondary component and its dependence on component parameters.
We derived LOSVDs for every spectrum using a set of regularization parameter $\lambda$ (30 logarithmically sampled steps from 0.01 to 10).
Then we decomposed the LOSVDs into two Gaussians and calculate the sum of the relative errors of the bestfit parameters $\sigma_1$, $V_2$ and $W_2$.
The minimum total error corresponds to the optimal $\lambda$, which usually coincides with the visual optimal identification.
Then we focused on the most prominent cases of counter-rotation, where $V_2 / \sigma_1 > 4$ and examined samples of synthetic spectra spanning a grid of age $\Delta$ Age = $\log{\text{Age}_1/\text{Age}_2}$ metallicity $\Delta$ Met = Met$_1$ - Met$_2$ differences from 0 to 1 dex.
At each grid point, the relative error $\partial$W$_2$ increased as W$_2$ decreased.
At each point of the grid, a straight line was drawn in the plane of relative errors $\partial$W$_2$ and W$_2$.
We adopted a minimum detectable weight W$_{min}$ at the relative error level of $\partial$W = 30\%.

Fig.~\ref{fig:age_met_weights} shows the minimum detectable weight of the secondary component as a function of the difference in age and metallicity of the components.
The following plane approximates the individual values in the space ($\Delta$ Met, $\Delta$ Age, W$_{min}$):
$$\text{W}_{min} = (0.101 \pm 0.016) + (0.040 \pm 0.032)\Delta~\text{Age} + (0.205 \pm 0.039) \Delta~\text{Met}$$ 
This suggests that the two-peak LOSVD shape can be recovered if the contribution of the secondary component is at least 10\%.
In addition, the difference in the age of the components has almost no effect on the result, while the difference in the metallicity is very important.

\section{Real-world application}

\begin{figure}
    \centering
    \includegraphics[width=0.55\columnwidth,viewport=20 900 930 1320,clip]{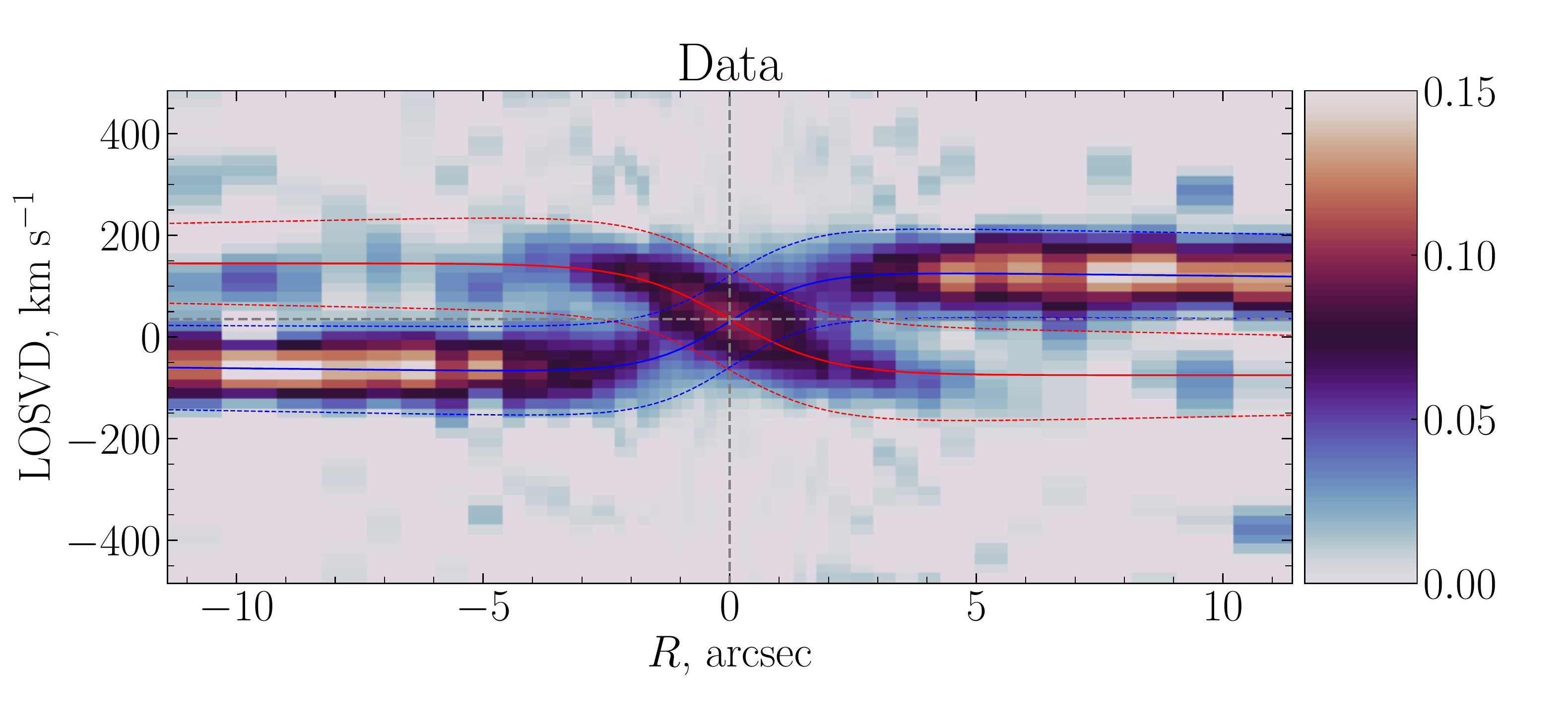}
    \includegraphics[width=0.42\columnwidth,viewport=0 680 480 1000,clip]{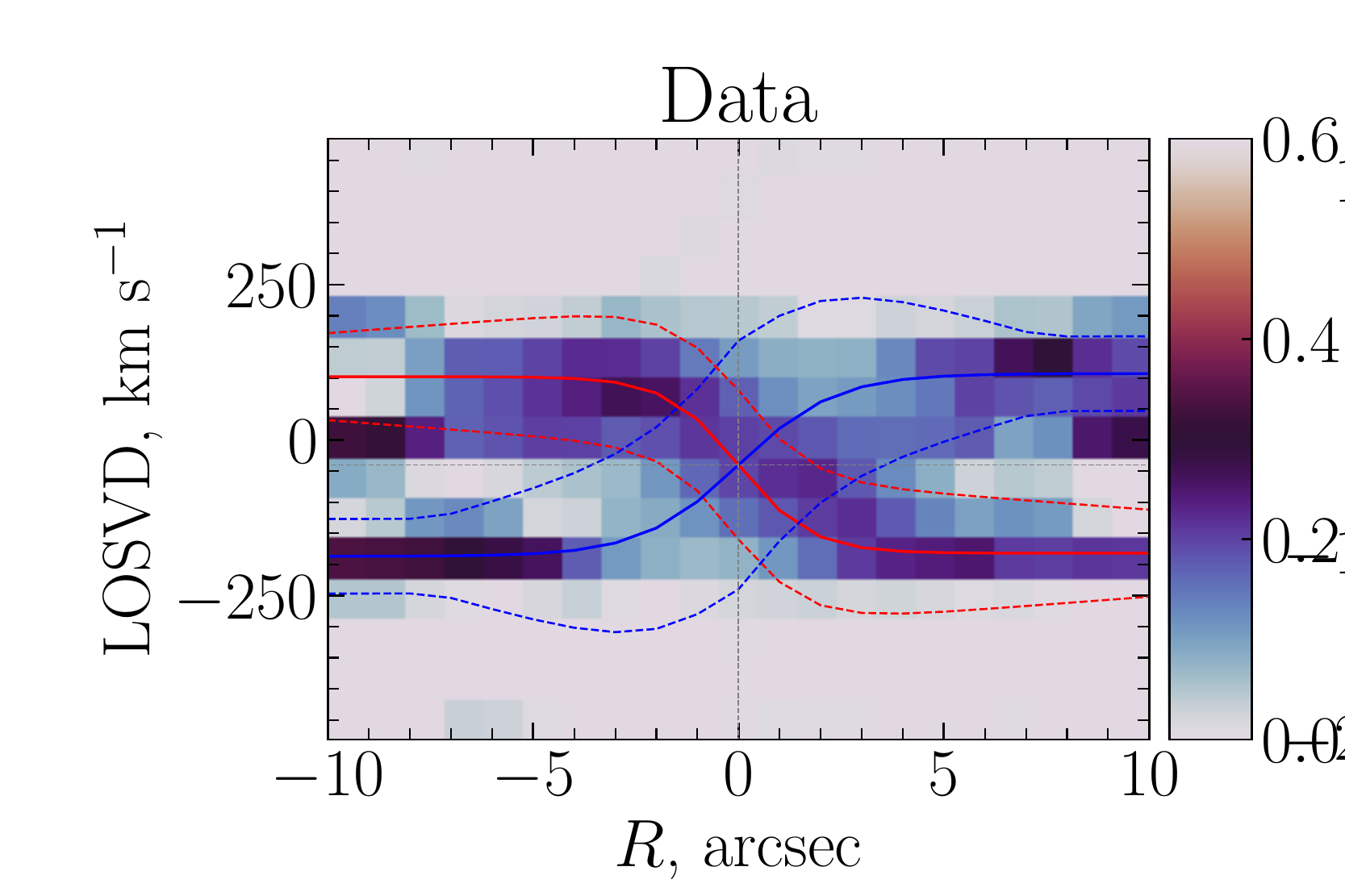}
    \caption{Examples of recovery of stellar LOSVDs in counter-rotating galaxies PGC 066551 based on the long-slit spectrum (R$\approx$3000) taken at SALT telescope and galaxy MaNGA 1-136248 based on IFU-spectrum MaNGA SDSS survey (R$\approx$2000).}
    \label{fig:losvds}
\end{figure}

We are applying the described technique to study a sample of counter-rotating galaxies identified in the MaNGA survey and followed up with  higher and deeper long-slit spectroscopy at the Russian BTA 6 telescope and the South African Large Telescope (SALT).
Fig.~\ref{fig:losvds} shows recovered stellar LOSVD for one of studied galaxy PGC~066551. 
At the moment we are mostly applying the described technique to obtain a reasonable initial guesses for the two-component parametrical \textsc{NBursts} analysis, which correctly accounts for possible differences in the properties of the stellar populations both disks. 



\textbf{Acknowledgments.} DG thanks the ADASS organizing committee for the provided financial aid. IK and DG are supported by the RScF grant 21-72-00036 and the Interdisciplinary Scientific and Educational School of Moscow University ``Fundamental and Applied Space Research''.

\bibliography{X4-002}

\begin{thebibliography}{}
\expandafter\ifx\csname natexlab\endcsname\relax\def\natexlab#1{#1}\fi
\expandafter\ifx\csname url\endcsname\relax
  \def\url#1{\texttt{#1}}\fi
\expandafter\ifx\csname urlprefix\endcsname\relax\def\urlprefix{URL }\fi
\providecommand{\eprint}[2][]{\url{#2}}

\bibitem[{{Cappellari}(2017)}]{Cappellari2017MNRAS.466..798C}
{Cappellari}, M. 2017, \mnras, 466, 798. \eprint{1607.08538}

\bibitem[{{Chilingarian} et~al.(2007){Chilingarian}, {Prugniel}, {Sil'Chenko},
  \& {Koleva}}]{Chilingarian2007IAUS..241..175C}
{Chilingarian}, I., {et~al.} 2007, in Stellar Populations as Building Blocks of
  Galaxies, edited by A.~{Vazdekis}, \& R.~{Peletier}, vol. 241, 175.
  \eprint{0709.3047}

\bibitem[{{Falc{\'o}n-Barroso} \& {Martig}(2021)}]{2021A&A...646A..31F}
{Falc{\'o}n-Barroso}, J., \& {Martig}, M. 2021, \aap, 646, A31.
  \eprint{2011.12023}

\bibitem[{{Kasparova} et~al.(2020){Kasparova}, {Katkov}, \&
  {Chilingarian}}]{2020MNRAS.493.5464K}
{Kasparova}, A.~V., {Katkov}, I.~Y., \& {Chilingarian}, I.~V. 2020, \mnras,
  493, 5464. \eprint{1912.04887}

\bibitem[{{Katkov} et~al.(2013){Katkov}, {Sil'chenko}, \&
  {Afanasiev}}]{Katkov2013ApJ...769..105K}
{Katkov}, I.~Y., {Sil'chenko}, O.~K., \& {Afanasiev}, V.~L. 2013, \apj, 769,
  105. \eprint{1304.3339}

\bibitem[{{Katkov} et~al.(2016){Katkov}, {Sil'chenko}, {Chilingarian},
  {Uklein}, \& {Egorov}}]{2016MNRAS.461.2068K}
{Katkov}, I.~Y., {et~al.} 2016, \mnras, 461, 2068. \eprint{1606.04862}

\bibitem[{{Ocvirk} et~al.(2006){Ocvirk}, {Pichon}, {Lan{\c{c}}on}, \&
  {Thi{\'e}baut}}]{2006MNRAS.365...74O}
{Ocvirk}, P., {et~al.} 2006, \mnras, 365, 74. \eprint{astro-ph/0507002}

\bibitem[{{Rix} \& {White}(1992)}]{1992MNRAS.254..389R}
{Rix}, H.-W., \& {White}, S. D.~M. 1992, \mnras, 254, 389

\bibitem[{{van der Marel} \& {Franx}(1993)}]{vanderMarel1993ApJ...407..525V}
{van der Marel}, R.~P., \& {Franx}, M. 1993, \apj, 407, 525

\end{thebibliography}


\end{document}